\documentclass[pre,reprint,groupaddress]{revtex4-2}
\usepackage{graphicx}
\usepackage{amssymb,amsmath}
\usepackage[utf8]{inputenc}
\usepackage[T1]{fontenc}
\usepackage[english]{babel}

\begin{document}
\title{Epidemic-Driven Collapse in a System with Limited Economic Resource. II}
\author{I.\,S.~Gandzha}
\email{gandzha@iop.kiev.ua}
\author{O.\,V.~Kliushnichenko}
\email{kliushnychenko@iop.kiev.ua}
\author{S.\,P.~Lukyanets}
\email{lukyan@iop.kiev.ua}
\affiliation{Institute of Physics, Nat.~Acad.~of~Sci.~of Ukraine, Prosp.~Nauky 46, Kyiv 03028, Ukraine}

\begin{abstract}
We consider a possibility of socioeconomic collapse caused by the spread of epidemic. To this end, we exploit a simple SIS-like (susceptible-infected-susceptible) model with negative feedback between the infected population size and a collective economic resource associated with the average amount of money or income per economic agent. The coupling mechanism in such a system is supposed to be of activation type, with the recovery rate governed by the Arrhenius-like law. In this case, economic resource formally plays the role of effective market temperature and the minimum level of resource consumption is associated with activation energy. Such a coupling can result in the collapsing effect opposite to thermal explosion, so that the epidemic could ultimately drive the system to a collapse at nonzero activation energy because of the limited resource. In this case, the system can no longer stabilize and return to the stable pre-epidemic state or a poorer post-epidemic state. We demonstrate that the system's collapse can partially be mitigated by external subsidies meaning constant resource inflow from some external source or by means of debt interpreted as a negative resource. We also consider a simple quarantine scenario and show that it can lead to different socioeconomic outcomes, depending on initial resource (market temperature) and the minimum level of resource consumption (activation energy).
\end{abstract}
\maketitle

\section{\label{intro}Introduction}

Systemic shocks like the outbreak of epidemics and contagion spreading inevitably lead to negative socioeconomic outcomes \cite{SystemicShock_2020}. A dramatic example is the spread of COVID-19 that had a domino effect on both the social and economic levels. Different countries and governments resorted to different mitigation strategies and quarantine measures~\cite{Anderson_Lancet_2020-03}. Countries with a higher resource level (economic or financial) could use stricter quarantine measures, while for countries with a lower resource the use of such measures led to the economic collapse, at least for a number of industries and/or social groups. The problem of strategy selection reduces to problems of optimal control theory for feedback systems \cite{Economics_2014,Optimization_SIS_2017,Percolation_2017} or to the theory of games in a more general case \cite{Bauch_Earn_2004}. The use of one or another action strategy reduces to the classical problem of choice \cite{Economics}, i.e., to the definition of the sacrifice, when the salvation of someone or something is only possible at the expense of the other one.

Population systems, as well as the economic ones, can exhibit critical behaviors or instabilities \cite{China_PR_2018,China_2018}. For population systems, such behaviors can be triggered by the onset and spreading of an epidemic, when many individuals become infected by some contagion \cite{Swiss_PRL_2017}. The dynamics of epidemic-like processes is determined by transitions between the inner states of individuals, such as susceptible, infected, ill, recovered, etc. It is usually described in terms of mean numbers of individuals in these states. For economic systems, critical behaviors can be triggered by the onset of a financial crisis or bankruptcy spreading \cite{Cascades_PRE_2015}, when many economic agents lose their money, income, or wealth. Such processes are described by another dynamical variable, usually associated with money, and are characterized by the distribution of money or income among economic agents. The equilibrium distribution of money or income in systems of economic agents was shown to obey the Boltzmann-Gibbs statistics \cite{Yakovenko_2000,Yakovenko_rmp_2009,Yakovenko_2010,Efthimiou_2016} or, in a more general form, Planck and Bose-like distributions~\cite{Kusmartsev_2011,Xu_epl_2015,Efthimiou_2016}. In this case, the average amount of money or income per agent is associated with the effective temperature, which is sometimes called the economic or market temperature \cite{Yakovenko_2000}. The crisis state in such systems is regarded, in particular, as the one corresponding to the Bose-Einstein condensate-like state~\cite{Kusmartsev_2011,Xu_epl_2015,Tao_2010,Rashkovskiy_2019}.

If economic agents can also be susceptible to some contagion and their state of being ill/healthy means the state of being passive/active from the economic viewpoint, then the following questions arise: (i) What is the effect of agent's transition from the active state to the passive one on the agent's income? and (ii) How does the passive-to-active transition rate depend on agent’s income or other economic resource?

In what follows, we will consider some selected group of economic agents as a particular example of a socioeconomic system. We identify the agent's resource with the amount of money or income the agent possesses. The agent's state is determined by the amount of agent’s resource and by two inner states---the active and passive ones.

In this context, a general problem of interest is to describe the dynamical response of a socioeconomic system being initially in equilibrium on an epidemic-like shock or, more precisely, on an applied ``external field'' such as contagion. This field results in transitions between two agent states (active and passive).
To be consistent, it would be desirable to exploit the kinetic approach based on the Boltzmann equation, where the collision mechanism of epidemic spreading \cite{hao} as well as the resource or money redistribution \cite{Yakovenko_rmp_2009,Chatterjee_2007,Chakra} could appropriately be taken into account. Usually, such non-equilibrium processes are described using a more simple approach based on evolution equations for the averaged variables \cite{BookSpringer2015,BookSpringer2019,Swiss_2015}, such as the mean numbers of active (susceptible) and passive (infected) individuals in population and average resource, which formally corresponds to the market temperature. As a matter of fact, collision mechanisms determine the form or symmetry of equations for the averaged variables and the corresponding rate constants, such as transmission rate, recovery rate, production rate, etc. To describe the socioeconomic interplay, we have to make some assumptions about the mechanism of coupling between the spreading process and resource.

The resource's influence on the spreading process can naturally be taken into account by means of the resource-dependent recovery rate \cite{Swiss_2015,China_PRE_2019}, which is one of the basic parameters in many spreading models \cite{RevModPhys_2015}. In particular, it can approximately be fitted from empirical data as a function of the ratio between the infected population size and the average amount of resource devoted to infected individuals \cite{China_PRE_2019}. On the other hand, the influence of the spreading process on resource can be taken into account in different ways, depending on the economic model adopted. Such an economic model can imply either the direct load on resource (depending on the infected population size) or indirect mechanisms like taxes, etc.

The simplest example of the epidemic-resource coupling has been demonstrated for the basic susceptible-infected-susceptible (SIS) epidemic model, where the recovery rate was set dependent on the resource (budget) availability \cite{Swiss_2015}. A sufficiently wide class of model coupling functions was introduced to take into account the influence of the budget on the recovery rate. The counter effect on the budget was of direct and almost reciprocal character described by the same model function. The epidemic was shown to spiral out of control into ``explosive'' spread if the cost of recovery was above some critical cost. The similar explosive epidemic spreading can be observed in the case of connectivity disruption in networks \cite{Swiss_PRE_2016}. The spread of concepts, memes, hashtags as well as online rumor cascades can also be explosive~\cite{PRL_2020}.

In this work, we consider a possibility for the coupling mechanism in the epidemic-resource system to be of activation type, with the recovery rate governed by the Arrhenius-like law. We show that such a coupling can lead to the system collapse caused by an epidemic shock.

The recovery rate is generally determined by the quality of provision with medical services and food, apart from the individual peculiarities of the given member of population. The quickest recovery depends on the cost of medical services and the bare subsistence level of consumption $E$, as well as on the availability of resource for the given economic agent. Since the cost of services is fixed, the service is terminated if there is no sufficient resource. In other words, the parameter $E$ serves as the height of some energy barrier peculiar to the given system. Therefore, we can suppose that the recovery rate, i.e., the mean frequency of transition from the agent's passive state to the active one, has an activation-type dependence, similar to the temperature dependence of common activation processes with activation energy~$E$ \cite{Laidler,Stiller}.

In physical systems, the  rate of over-barrier transitions is usually determined by the Arrhenius exponential factor $\exp(-E/k_B T)$ resulting from the Boltzmann statistics \cite{Stiller}, where $T$ is temperature and $k_B$ is Boltzmann constant. The transition rate $\gamma$ can roughly be estimated using an equilibrium particle distribution function $f(\epsilon)$,
$$
\gamma\approx\int_E^\infty g(\epsilon)f(\epsilon)d\epsilon,
$$
where the factor $g(\epsilon)$ takes into account a dependence of the elementary transition probability, attempt frequency, density of states, etc. on particle energy $\epsilon$. When this dependence is weak, we have $\gamma\approx \gamma_0\int_E^\infty f(\epsilon)d\epsilon$. In the case of Boltzmann statistics, this integral (which identifies the probability that the particle energy is greater than the activation energy $E$) gives the Arrhenius exponential factor.

On the other hand, as was shown by Dr\u{a}gulescu and Yakovenko \cite{Yakovenko_2000}, the equilibrium distribution of income or money $m$ for single economic agents is also governed by the exponential Boltzmann law, $f(m)=T^{-1}\exp(-m/T)$, at least for low and middle income classes \cite{Yakovenko_rmp_2009,Yakovenko_2010,Yakovenko_2019}. Here the effective temperature $T=\langle m\rangle$ is associated with the average amount of money or average income per economic agent.

The above points enable us to rewrite the recovery rate in the Arrhenius-like form, namely
\begin{equation}\label{eq:gamma}
\gamma(\rho)=\gamma_0 \,\exp(-E/\rho),
\end{equation}
where resource $\rho\equiv T$ is the average amount of money or average income per economic agent. Although the factor $\gamma_0$ generally depends on $\rho$, we set it to be constant for the sake of simplicity. The activation energy $E$ corresponds to the minimum level of resource consumption. By analogy to the cumulative distribution of purchasing power introduced in Ref.~\cite{Yakovenko_2000}, the recovery rate given by Eq.~(\ref{eq:gamma}) can formally be regarded as the cumulative distribution function of resource consumption associated with the fight against the epidemic.

The activation process described by the Arrhenius-like law (\ref{eq:gamma}) implies that the system can exhibit the so-called explosive (or catastrophic) instability \cite{PRL_1966}. For example, when a chemical reaction occurs with the release of heat and has an activation character, it goes faster at higher temperatures. This leads to yet greater temperature increase and ultimately to a thermal explosion, which is described in the framework of the Zel'dovich-Frank-Kamenetskii theory \cite{Zeldovich_Frank,Smirnov,Novozh_2018}.

The fight against the epidemic involves similar catastrophic processes. The spread of epidemic and the associated quarantine measures result in the reduction of the collective resource $\rho$. When resource is depleted, the quality of medical services drops and the recovery rate goes down. As a result, the number of active members in population decreases. This, in turn, leads to a further reduction of the collective resource, with the level of income needed for the basic survival being lower and lower. Such a scenario finally results in the ultimate collapse of the system---the effect opposite to thermal explosion.

To illustrate the epidemic-driven collapse dynamics, we resort to the simplest SIS-like model supplemented with resource equation, supposing that the epidemic-resource coupling is of activation type [Eq.~(\ref{eq:gamma})]. This model describes the temporal coevolution of the mean number of active agents and the average resource or market temperature for some selected social group.

\section{\label{model} Basic equations for an epidemic-resource system}

Our model is based on the simplest SIS model (as in Refs.~\cite{China_PR_2018,Swiss_PRL_2017,Swiss_2015}), where a certain group of individuals or economic agents is divided into two subgroups: active and passive ones in the economic sense. Active agents can be infected with some contagion and pass into the passive state being, e.g., ill. In terms of the classical SIS model, active agents correspond to the susceptible ones, and passive agents correspond to the infected ones.

Susceptible individuals are infected at some transmission rate $\beta$, which is defined as a product of the contact rate and the probability that a contact of an infected individual with a susceptible individual results in transmission. Infected (passive) individuals recover and become susceptible (active) again with recovery rate $\gamma$ given by relation (\ref{eq:gamma}) implying that the recovery process is governed by the general economic situation characterized by the average resource $\rho$ (market temperature) and activation parameter $E$, which identifies the minimum level of resource consumption. The corresponding mathematical model is given by the following two ODEs:
\begin{subequations}\label{eq:SIS}
\begin{align}
\partial_t s &= - \beta\, s\,\bigl(1-s\bigr) + \gamma(\rho)\,\bigl(1-s\bigr),\label{eq:SISa}\\
\partial_{t} \rho &= G\,s-\Gamma\rho+\Lambda.\label{eq:SISb}
\end{align}
\end{subequations}
The operator $\partial_t$ stands for the derivative with respect to time $t$. Here $s$ is the number density of susceptible individuals (active agents) and $i=1-s$ is the number density of infected individuals (passive agents). The total number of susceptible and infected individuals is assumed to be constant.

The function $\rho$ represents the average resource associated with the average amount of money or income per economic agent. The acquisition of this resource per unit time is proportional to the number density of active (working) agents, $s$. The acquisition rate $G$ formalizes the resource amount acquired by them per unit time. Note that here, in line with the comments made in Refs.~\cite{Yakovenko_2000,Yakovenko_rmp_2009,Yakovenko_2010}, we suppose that money cannot actually be produced (the total money balance remains constant) and, therefore, use the term ``resource acquisition rate'' rather than ``resource production rate.'' The former term was used, in particular, in Ref.~\cite{Amado_2019} to describe a rate at which the agents acquire resources from the environment in a stochastic resource-based model.

The second term, $\Gamma\,\rho$, formally describes the collective expenses or taxes. Roughly speaking, the expenses are assumed to be proportional to earnings. Thus, the coefficient $\Gamma$ represents the resource consumption rate.

The parameter $\Lambda$ represents a resource source (constant resource inflow into the system from some external reservoir) or a resource sink (constant resource outflow from the system). When $\Lambda>0$, resource is fed into the system (e.g., in the form of subsidies) from some external source, e.g., a central bank or central government. In this case, our resource balance equation is in accord with an equation for the budget constraint in the macroeconomic model considered in Ref.~\cite{Macroeconomic_2020}. A similar equation for the resource dynamics with resource consumption and constant resource inflow was analyzed in Ref.~\cite{Sugiarto_PRL_2017} in application to a system of interacting agents that exploit a common pool resource. When $\Lambda<0$, resource flows out from the system, e.g., in the form of infrastructure expenses, depreciation, rent, interest payments, or other fixed expenses that do not depend on the agent's state. The case of negative $\Lambda$ was also discussed in Ref.~\cite{Swiss_2015}, but in contrast to the budget equation considered therein, we consider the indirect influence of the epidemic on resource via taxes, collective expenses, or infrastructure. Finally, a resource exchange model with negative $\Lambda$ describing interest returns was considered in Ref.~\cite{Yakovenko_2000}.

It should be noted that our group of agents is an open subsystem embedded in a broader system. As a matter of fact, Eq.~(\ref{eq:SISb}) describes the average balance of money redistribution between our subsystem and the environment (external reservoir). In this subsystem, money is not produced (e.g., by printing) but is supplied from the external subsystem. In general, the term $\Lambda$ should also depend on dynamical variables, in particular on $s$ and $i$, but this average flow is determined by the total number of agents in the group, which is constant, $s+i=1$. As a result, $\Lambda$ behaves as a permanent external source or sink in Eq.~(\ref{eq:SISb}). This equation formally corresponds to the equation of the average heat balance between the system and its environment.

Our point of interest is the dynamical response of the socioeconomic system described by Eqs.~(\ref{eq:SIS}) and being initially in the equilibrium state $s_0=1$ and $\rho_0=$ \mbox{$(G+\Lambda)/\Gamma$} on an epidemic-like shock caused by the emergence of some contagion.
In the framework of our simple model, a formal field associated with contagion is absent. In general, eliminating a field that interacts with particles results in an induced particle-particle interaction and a memory effect (see, e.g., Ref.~\cite{Gardiner}). In our case, this induced interaction between the agents affects the agent's inner state and is formally represented by the phenomenological collision term $\propto s(1-s)$ in Eq.~(\ref{eq:SISa}), where any possible agent-agent correlations and memory effects are absent. It is the interaction that drives the system out of equilibrium. In this regard, the system's response on the contagion field can be considered in terms of system's response on a perturbation of initial conditions at $t = 0$,
\begin{equation}\label{eq:initial}
s(0)=s_0-\delta s(0)=1-\delta s(0),\quad \rho(0) = \rho_0,
\end{equation}
where $\delta s(0)=i(0)$ is the initial number density of infected individuals. It should be noted that since Eqs.~(\ref{eq:SIS}) admit not only the parametric instability but also the instability with respect to initial conditions, we shall also take into consideration a possible initial resource perturbation $\rho(0)=\rho_0-\delta\rho(0)$.

In the case of unlimited resource ($E\ll\rho$), Eq.~(\ref{eq:SISa}) for $s$ reduces to the basic SIS model, whose solutions are well studied~\cite{BookSpringer2015,BookSpringer2019}. This model is due to Kermack and McKendrick \cite{KermackMcKendrick}; it is also known as the Schl\"{o}gl~I model describing autocatalytic chemical reactions \cite{Schlogl_1972}. The SIS model can also be derived as the mean-field approximation to more general network models \cite{RevModPhys_2015,Swiss_PRL_2017}.

Let us investigate the effect of nonzero activation parameter $E$ (activation energy) on the coupled epidemic-resource dynamics described by system~(\ref{eq:SIS}). We first consider the simplest case with no resource inflow or outflow ($\Lambda=0$). The case of nonzero $\Lambda$ is considered afterwards.

\section{Analysis}

\subsection{No resource inflow or outflow ($\Lambda=0$)}

We first consider the simplest case with no resource inflow or outflow ($\Lambda=0$). As we understand, if there are no systemic shocks like an epidemic, there exists a stationary equilibrium state with $\rho = \rho_0 = \mathrm{const}$. It is called the disease-free equilibrium and is given by a trivial stationary solution to Eqs.~(\ref{eq:SIS}), namely, $s_0=1$, $\rho_0=G/\,\Gamma$.

In the case of epidemic, the system can go out from the disease-free equilibrium, with resource decreasing. Indeed, apart from the disease-free trivial solution $\rho_0$, another stationary solution to the equation for $\rho$ is given by $\rho^{*}=G\,s^{*}/\,\Gamma$, where $s^{*}$ is given by the following transcendental equation:
\begin{equation}\label{eq:SIS_trans}
s^{*}\,\ln\left(\mathcal{R}_0 s^{*}\right)=-\mathcal{E},
\end{equation}
provided that $s^{*}>0$. The dimensionless parameter $\mathcal{R}_0=\beta/\,\gamma_0$ is well known as the basic reproduction number. It defines the average number of transmissions one infected individual makes in the entire susceptible compartment during the entire time of being infected. The dimensionless parameter $\mathcal{E}=E/\,\rho_0$ is the activation energy normalized by the stationary resource value $\rho_0$.

\begin{figure}[t]
\includegraphics[width=0.75\columnwidth]{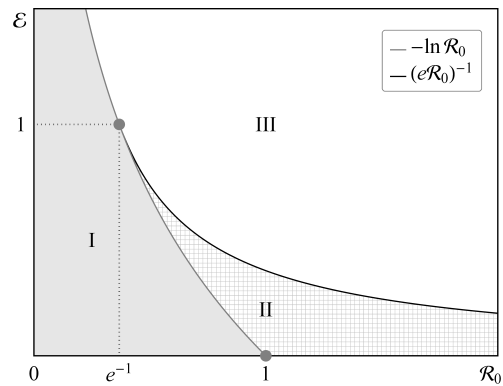}
\caption{\label{fig:phase}Phase diagram for the coupled epidemic-resource system described by Eqs.~(\ref{eq:SIS}) with $\Lambda=0$. Three states (phases) are possible: (I) disease-free equilibrium, (II) endemic equilibrium, and (III) collapse. Phase (I) and phase~(II) coexist with phase (III). When $\mathcal{E}>\mathcal{E}_b$ (with $\mathcal{E}_b$ given by Eq.~(\ref{eq:Eb}) and $\mathcal{E}=E/\,\rho_0$), the system collapses at any initial conditions $\{s(0),\,\rho(0)\}$. When $\mathcal{E}_e<\mathcal{E}\leqslant\mathcal{E}_c$ and $\mathcal{R}_0\geqslant e^{-1}$ (with $\mathcal{E}_c$ and $\mathcal{E}_e$ given by Eqs. (\ref{eq:E_c}) and (\ref{eq:E_log}), respectively), the system evolves to the endemic state at sufficiently small perturbations of  equilibrium initial conditions, $\delta s(0)\ll 1$ and $\delta \rho(0)\ll 1$ [see Eq.~(\ref{eq:initial})], or collapses if these perturbations are large enough. When $\mathcal{E}\leqslant\mathcal{E}_e$, the system returns to the disease-free equilibrium at sufficiently small perturbations of equilibrium initial conditions or otherwise collapses.}
\end{figure}

\begin{figure*}[!]
\includegraphics[width=\textwidth]{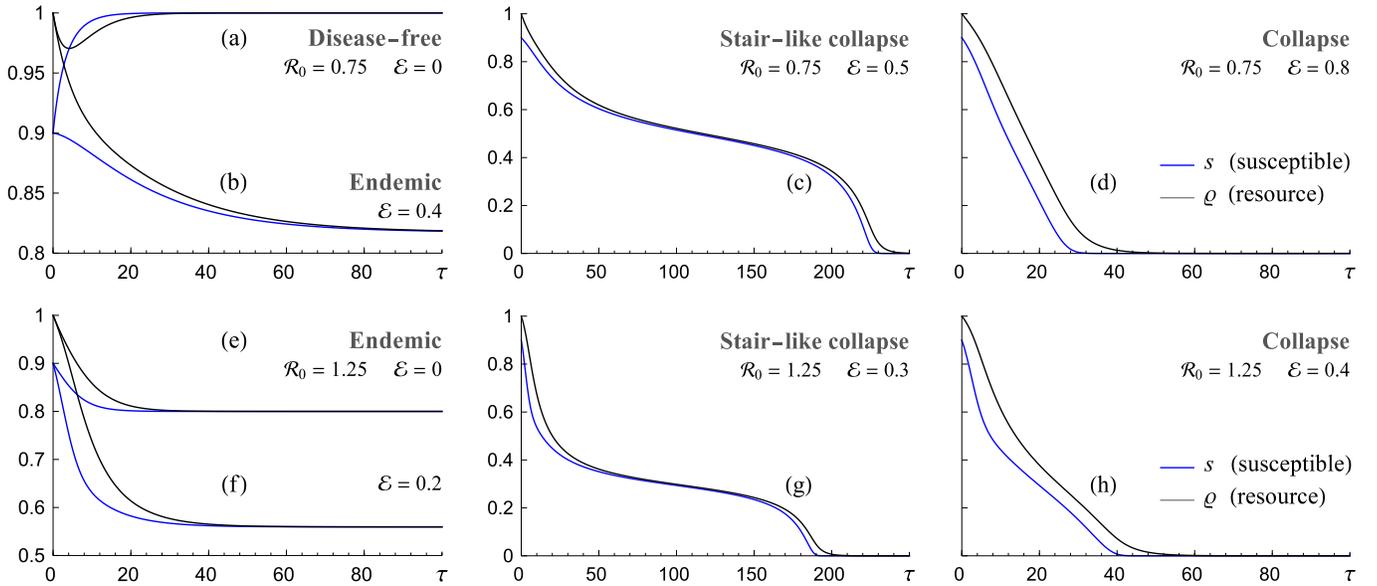}
\caption{\label{fig:SIS1}(Color online) The number density of susceptible (active) individuals and normalized resource function $\varrho=\rho/\rho_0$ versus dimensionless time $\tau = \gamma_0\,t$ in the coupled epidemic-resource system described by Eqs.~(\ref{eq:SIS}) with $\Gamma/\gamma_0=0.2$, $\mathcal{R}_0=0.75$ or $\mathcal{R}_0=1.25$, and various normalized activation energies $\mathcal{E}$ in the case of no resource inflow or outflow ($\Lambda=0$). The initial conditions are $i(0)=0.1$, $\varrho(0) = 1$. When $\mathcal{E}=0$, the system evolves to (a)~disease-free equilibrium at $\mathcal{R}_0<1$ and (e) endemic equilibrium at $\mathcal{R}_0>1$. When $\mathcal{E}>0$, the system evolves either to the endemic equilibrium (phase II in Fig.~\ref{fig:phase}) both for (b)~$\mathcal{R}_0<1$ and (f) $\mathcal{R}_0>1$ or collapses (phase~III in Fig.~\ref{fig:phase}). When $\mathcal{E}$ is above the critical value $\mathcal{E}_c$ given by Eq.~(\ref{eq:E_c}) but still close to it, the system first tries to occupy the quasi-stationary endemic state (which no longer exists). This process can take quite a long time and then the system finally collapses both for (c) $\mathcal{R}_0<1$ and (g) $\mathcal{R}_0>1$ (the so-called stair-like collapse). At larger activation energies, the collapse is very fast with no intermediate quasi-stationary evolution both for (d) $\mathcal{R}_0<1$ and (h)~$\mathcal{R}_0>1$.}
\end{figure*}

When $\mathcal{E}=0$, Eq.~(\ref{eq:SIS_trans}) has one solution, $s^{*}=\mathcal{R}_0^{-1}$ (since $s^{*}\ne0$). It is stable at $\mathcal{R}_0 > 1$ and defines the endemic equilibrium point. Thus, as it is well-known for the case of unlimited resource $\mathcal{E}=0$, the disease-free equilibrium is stable when $\mathcal{R}_0\leqslant 1$, and there is no epidemic outbreak \cite{BookSpringer2015,BookSpringer2019}. When $\mathcal{R}_0 > 1$, the disease-free equilibrium is unstable, and the system evolves to the new equilibrium state $\{s^*,\rho^*\}$ called the endemic equilibrium.

When $\mathcal{E}>0$, there are several possible cases. For $0<\mathcal{E}<\mathcal{E}_c$ and $\mathcal{R}_0\geqslant e^{-1}$, with
\begin{equation}\label{eq:E_c}
\mathcal{E}_c=\left(e\mathcal{R}_0\right)^{-1},
\end{equation}
Eq.~(\ref{eq:SIS_trans}) has two solutions: $s^{*}_1 > \mathcal{E}_c$ (which defines the endemic equilibrium point) and $0<s^{*}_2 < \mathcal{E}_c$ (which is always unstable). For $\mathcal{E}=\mathcal{E}_c$, there is one solution $s^{*}_{1,2} = \mathcal{E}_c$. Finally, there are no real solutions for $\mathcal{E}>\mathcal{E}_c$.

Except for the condition $0<\mathcal{E}\leqslant\mathcal{E}_c$, the endemic equilibrium point should also meet the requirement of $s^{*}< 1$, which effectively implies that $\mathcal{E}>\mathcal{E}_e$, where
\begin{equation}\label{eq:E_log}
\mathcal{E}_e=-\ln\mathcal{R}_0.
\end{equation}
The same relation can be obtained from the stability analysis of the disease-free stationary solution $s_0=1$. The  disease-free equilibrium is stable at $\mathcal{E}\leqslant \mathcal{E}_e$ and unstable at $\mathcal{E} > \mathcal{E}_e$.

In the case $\mathcal{E}>0$, system~(\ref{eq:SIS}) also possesses another stable stationary solution given by $s^{*}_c=0$, $\rho^{*}_c\rightarrow0$. This means that at any nonzero $\mathcal{E}$ and $\mathcal{R}_0$ there exist such initial conditions $\{s(0),\,\rho(0)\}$ that the system collapses (\mbox{$s\rightarrow 0$}) because of resource depletion ($\rho\rightarrow0$). When $\mathcal{E}\leqslant \mathcal{E}_e$, the stationary point $s^{*}_c=0$ coexists with the disease-free equilibrium $s_0=1$. The system evolves to one of these two points, depending on initial conditions $\{s(0),\,\rho(0)\}$. Similarly, the stationary point $s^{*}_c=0$ coexists with the endemic equilibrium $s^{*}$ when $\mathcal{E}_e<\mathcal{E}\leqslant \mathcal{E}_c$ and $\mathcal{R}_0\geqslant e^{-1}$. For all other $\mathcal{E}$ and $\mathcal{R}_0$, the system collapses at any initial conditions.

Thus, relations (\ref{eq:E_c}) and (\ref{eq:E_log}) define two critical curves $\mathcal{E}_c(\mathcal{R}_0)$ and $\mathcal{E}_e(\mathcal{R}_0)$ in the $(\mathcal{R}_0,\,\mathcal{E})$ plane which determine the evolution scenario for dynamical system (\ref{eq:SIS}). Depending on the values of parameters $\mathcal{R}_0$ and $\mathcal{E}$, the system can evolve into three possible states (phases): (I) disease-free equilibrium, (II) endemic equilibrium, or (III) collapse. Figure \ref{fig:phase} shows the corresponding phase diagram. The bottom of the collapse domain is defined by a curve
\begin{equation}\label{eq:Eb}
\mathcal{E}_b = \left\{\begin{array}{l}
\mathcal{E}_c(\mathcal{R}_0),\; \mathcal{R}_0\geqslant e^{-1},\\
\mathcal{E}_e(\mathcal{R}_0),\; \mathcal{R}_0 < e^{-1}.
 \end{array}\right.
\end{equation}
The point of contact $A=(e^{-1},\,1)$ between the curves $\mathcal{E}_c(\mathcal{R}_0)$ and $\mathcal{E}_e(\mathcal{R}_0)$ defines a triple point between phases (I), (II), and (III).

The above analysis is supported by the results of numerical integration of Eqs.~(\ref{eq:SIS}) demonstrated in Fig.~\ref{fig:SIS1}.

When $\mathcal{E}=0$, the dynamics of system~(\ref{eq:SIS}) follows the basic SIS model. It evolves to the state of disease-free equilibrium at $\mathcal{R}_0\leqslant1$ [Fig.~\ref{fig:SIS1}(a)] and to the state of endemic equilibrium at $\mathcal{R}_0>1$ [Fig.~\ref{fig:SIS1}(e)].

When $\mathcal{E}>0$, some part of resource is consumed, and the number of susceptible (active) individuals decreases [Fig.~\ref{fig:SIS1}(f)]. There is a critical value $\mathcal{E}_e$ defined by formula~(\ref{eq:E_log}) at which the system evolves to the endemic equilibrium even at $\mathcal{R}_0<1$ [Fig.~\ref{fig:SIS1}(b)]. This scenario is impossible in the basic SIS model. For the activation energies larger than the critical value $\mathcal{E}_c$ defined by formula~(\ref{eq:E_c}), the endemic equilibrium is no longer stable and the system collapses to the state $s^{*}_c=0$, $\rho^{*}_c\rightarrow0$ [Fig.~\ref{fig:SIS1}(c,d)]. This means that all the individuals become infected (passive) and there is no resource to reverse the epidemic back. The same scenario is observed in the case $\mathcal{R}_0>1$ [Fig.~\ref{fig:SIS1}(g,h)]. Note that the endemic equilibrium shown in Fig.~\ref{fig:SIS1}(b,f) coexists with the collapse point, the system's dynamics switching from endemic to collapse at small $s(0)$ and $\rho(0)$.

When $\mathcal{E}$ is above the critical value $\mathcal{E}_c$ but still close to it, the system first tries to occupy the quasi-stationary endemic state (which no longer exists). This process can take quite a long time and then the system finally collapses [Fig.~\ref{fig:SIS1}(c,g)]. It resembles the well-known ``devil's staircase'' pattern \cite{Staircase}. At larger activation energies, the collapse is very fast with no intermediate quasi-stationary evolution [Fig.~\ref{fig:SIS1}(d,h)].

Thus, the system collapses at sufficiently high $\mathcal{R}_0=\beta/\,\gamma_0$ (high transmission rate $\beta$ or low recovery rate constant $\gamma_0$) and/or sufficiently high $\mathcal{E}=E/\,\rho_0$ (high activation energy $E$ or low initial equilibrium resource $\rho_0$). Otherwise, the system occupies either the disease-free equilibrium or endemic equilibrium. An additional resource deficit (large initial resource fluctuations $\delta\rho(0)$) can also lead to a collapse.

\subsection{Constant resource inflow ($\Lambda>0$)}

\begin{figure*}[!]
\includegraphics[width=\textwidth]{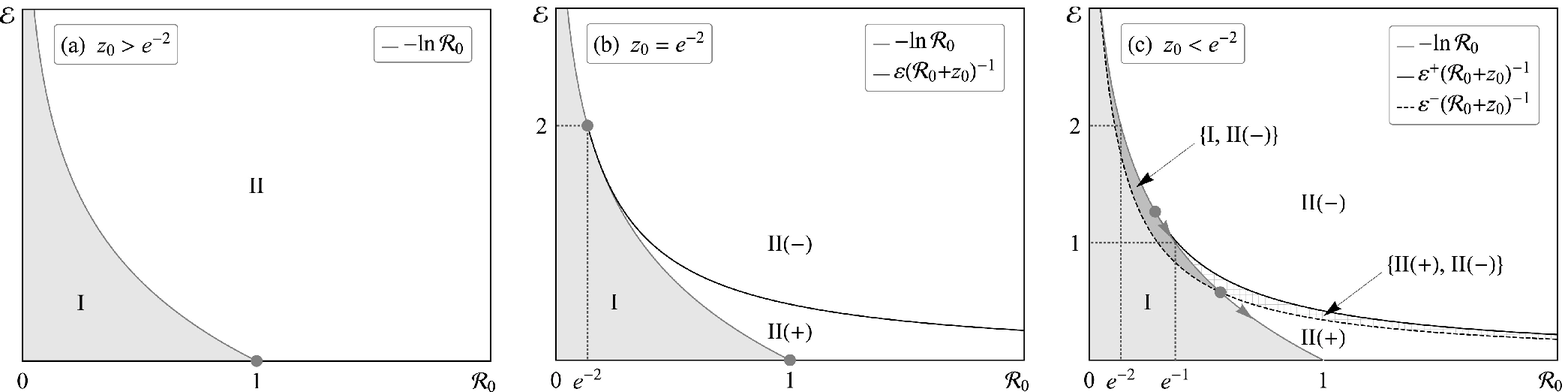}
\caption{\label{fig:phase_plus}Phase diagrams for the coupled epidemic-resource system described by Eqs.~(\ref{eq:SIS}) with $\Lambda>0$. Three cases are possible, depending on parameter $z_0 = s_\Lambda\mathcal{R}_0$: (a) the system can occupy either (I) disease-free equilibrium or (II) endemic equilibrium, depending on parameters $\mathcal{E}$ and $\mathcal{R}_0$; (b) critical point at which two distinct endemic phases II(+) and II(--) are formed out from the uniform endemic phase II; (c) the area of the endemic phase II(--) enlarges, with two new zones formed where phase II(--) coexists with phase (I) and phase II(+). The endemic phase II(--) turns into the collapse phase III (Fig.~\ref{fig:phase}) in the limit $\Lambda\rightarrow0$ ($z_0\rightarrow0$).}
\end{figure*}

\begin{figure*}[!]
\includegraphics[width=\textwidth]{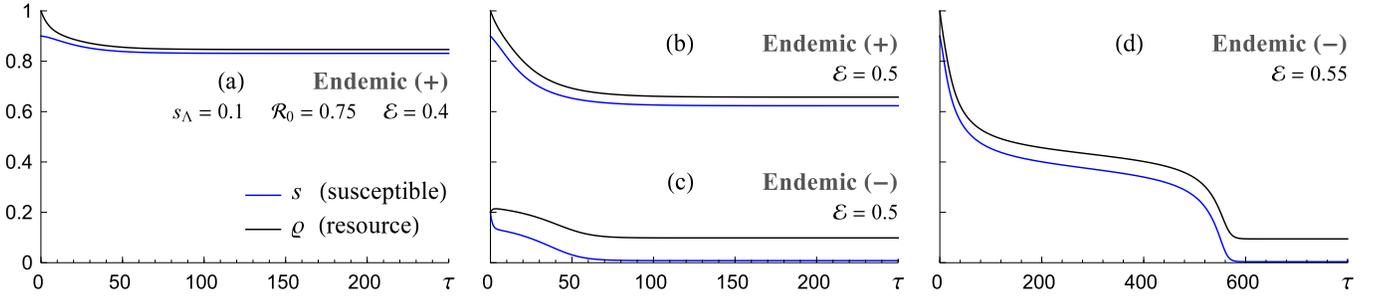}
\caption{\label{fig:SIS3}(Color online) The number density of susceptible (active) individuals and normalized resource function $\varrho=\rho/\rho_0$, with $\rho_0=(G+\Lambda)/\,\Gamma$, versus dimensionless time $\tau = \gamma_0\,t$ in the coupled epidemic-resource system described by Eqs.~(\ref{eq:SIS}) with $\Gamma/\gamma_0=0.2$ and $\mathcal{R}_0=0.75$ in the case of constant resource inflow ($\Lambda>0$, $s_\Lambda=0.1$). When (a) $\mathcal{E}=0.4$, the system evolves to the endemic state II(+) [see phase diagram in Fig.~\ref{fig:phase_plus}(c)]. When $\mathcal{E}=0.5$, the system evolves to one of the endemic states (b)~II(+) or (c)~II(--), depending on the initial conditions $\{s(0),\varrho(0)\}$. The same parameter values resulted in the collapse of the system in the case $\Lambda=0$ [Fig.~\ref{fig:SIS1}(c)], which is now mitigated owing to constant resource inflow. When (d) $\mathcal{E}=0.55$, the system evolves to the endemic state II(--).}
\end{figure*}

Here we will show that the ultimate collapse considered in the previous subsection is not possible when $\Lambda>0$. On the other hand, there can coexist several distinct endemic phases that turn into the single endemic phase and the collapse phase in the limit $\Lambda\rightarrow 0$.

Positive $\Lambda$ formally means constant resource inflow (e.g., in the form of subsidies) into the system from some external reservoir, e.g., a central bank or central government. This process is analogous to an influx of energy into a thermodynamical system from external sources. Nonzero $\Lambda$ breaks the symmetry of the equation for $\rho$, so that the collapse point $s^{*}_c=0$, $\rho^{*}_c\rightarrow0$ might no longer be its stationary solution. Indeed, the stationary solution to this equation at $\Lambda>0$ is
\begin{equation}\label{eq:rhostar_positive}
\rho^{*}= 
\frac{G\left(s^{*}+s_\Lambda\right)}{\Gamma},
\end{equation}
with $\rho^{*}>0$ for any $s^{*}\geqslant0$. The parameter $s_\Lambda = \Lambda/\,G$ defines the number density of active agents that would need to stay working to generate the amount of resource that is fed to the system from the external reservoir in the form of subsidies.

The stationary number density $s^{*}$ of active agents is given by the transcendental equation
\begin{equation}\label{eq:SIS_trans_Lambda_positive}
(s^{*}+s_\Lambda)\,\ln\left(\mathcal{R}_0 s^{*}\right)=-\mathcal{E}\left(1+s_\Lambda\right),
\end{equation}
where $\mathcal{E}=E/\,\rho_0$ and $\rho_0=(G+\Lambda)/\,\Gamma$ is the equilibrium disease-free resource value in the case when there is no epidemic ($s_0=1$). In contrast to Eq.~(\ref{eq:SIS_trans}), this equation always has one or several solutions for any $\mathcal{E}\geqslant0$ and $s_\Lambda > 0$.

When $s_\Lambda > s_\Lambda^{(c)}$, with
\begin{equation}
s_\Lambda^{(c)}=(e^{2}\mathcal{R}_0)^{-1},
\end{equation}
Eq.~(\ref{eq:SIS_trans_Lambda_positive}) always has one solution. This solution is stable for $\mathcal{E} > \mathcal{E}_e$, with the critical point $\mathcal{E}_e$ given by formula~(\ref{eq:E_log}). Thus, the system evolves to the disease-free equilibrium $\{s_0,\,\rho_0\}$ (phase~I) at $\mathcal{E} \leqslant \mathcal{E}_e$ and to the state of endemic equilibrium $\{s^*,\,\rho^*\}$ (phase~II) at $\mathcal{E} > \mathcal{E}_e$ for any initial conditions $\{s(0),\,\rho(0)\}$. The corresponding phase diagram is shown in Fig.~\ref{fig:phase_plus}(a).

When $s_\Lambda < s_\Lambda^{(c)}$, Eq.~(\ref{eq:SIS_trans_Lambda_positive}) has one, two, or three solutions, depending on $\mathcal{E}$ and $\mathcal{R}_0$. At sufficiently small $\mathcal{E}$, there is only one solution, $s^*_+$, which is close to $s^*=\mathcal{R}_0^{-1}$. As $\mathcal{E}$ increases and passes through a critical point
$$
\mathcal{E}_c^{-}=\varepsilon^{-}\bigl(\mathcal{R}_0(1+s_\Lambda)\bigr)^{-1},
$$
$\varepsilon^{-}$ being the absolute value of the local maximum of the transcendental function of $\mathcal{R}_0s^*$ in the left-hand side of Eq.~(\ref{eq:SIS_trans_Lambda_positive}), two other solutions (stable one, $s^*_{-}$, and unstable one, $s^*_{\pm}$) are born through a saddle-node bifurcation. At $\mathcal{E}=\mathcal{E}_c^{-}$, these two solutions (stable and unstable) coincide, and Eq.~(\ref{eq:SIS_trans_Lambda_positive}) has only two distinct solutions, $s^*_+$ and $s^*_{-}$. As $\mathcal{E}$ increases further, the unstable point $s^*_{\pm}$ and the stable point $s^*_+$ approach each other and finally collide and annihilate at $\mathcal{E}=\mathcal{E}_c^{+}$, with
$$
\mathcal{E}_c^{+}=\varepsilon^{+}\bigl(\mathcal{R}_0(1+s_\Lambda)\bigr)^{-1}
$$
and $\varepsilon^{+}$ being the absolute value of the local minimum of the transcendental function of $\mathcal{R}_0s^*$ in the left-hand side of Eq.~(\ref{eq:SIS_trans_Lambda_positive}). At $\mathcal{E}>\mathcal{E}_c^{+}$, Eq.~(\ref{eq:SIS_trans_Lambda_positive}) has only one solution, $s^*_{-}$, which is stable and tends to zero at large $\mathcal{E}$, but always remains finite. The corresponding phase diagram is shown in Fig.~\ref{fig:phase_plus}(c). The domains of disease-free equilibrium (I) and endemic equilibrium (II) are separated by the phase boundary $\mathcal{E} = \mathcal{E}_e$, as in the case of $\Lambda=0$. The endemic phase has three distinct zones: II(+) with the stable point $s^*_+$, II(--) with the stable point $s^*_-$, and the middle zone where the both stable points coexist. The similar zone of coexisting phases, (I) and II(--), has formed between the domains of disease-free equilibrium (I) and endemic equilibrium II(--).

Thus, the point $s_\Lambda = s_\Lambda^{(c)}$ [Fig.~\ref{fig:phase_plus}(b)] defines a critical $\Lambda$ at which two distinct endemic phases II(+) and II(--) are formed out from the uniform endemic phase II existing at $s_\Lambda > s_\Lambda^{(c)}$. When $s_\Lambda < s_\Lambda^{(c)}$, these two phases coexist at $\mathcal{E}_c^{-}<\mathcal{E}<\mathcal{E}_c^{+}$. As $s_\Lambda$ decreases from $s_\Lambda^{(c)}$ to zero, the phase boundaries $\mathcal{E}=\mathcal{E}_c^{+}$ and $\mathcal{E}=\mathcal{E}_c^{-}$ slide down across the logarithmic curve $\mathcal{E}=\mathcal{E}_e$ until they reach the boundaries $\mathcal{E}=\mathcal{E}_c$ [see formula (\ref{eq:E_c})] and $\mathcal{E}=0$, respectively. In this limiting case ($\Lambda = 0$), the phase diagram shown in Fig.~\ref{fig:phase_plus}(c) transforms into the phase diagram shown in Fig.~\ref{fig:phase}, with the endemic phase II(--) turning into the collapse phase III.

Figure \ref{fig:SIS3} demonstrates some results of numerical integration of Eqs.~(\ref{eq:SIS}) supporting the above conclusions for one particular case of $s_\Lambda < s_\Lambda^{(c)}$ and $\mathcal{R}_0<1$. When $\mathcal{E}_e<\mathcal{E}<\mathcal{E}_c^{-}$, the system evolves to the endemic state II(+) (stable point $s^*_+$)  at any initial conditions [Fig.~\ref{fig:SIS3}(a)]. When $\mathcal{E}_c^{-}<\mathcal{E}<\mathcal{E}_c^{+}$, two distinct endemic states, II(+) and II(--), are possible. The system evolves to one of them (stable points $s^*_+$ or $s^*_-$), depending on the initial conditions $\{s(0),\,\rho(0)\}$ [Fig.~\ref{fig:SIS3}(b,c)]. Note that the same parameter values ($\mathcal{E}$ and $\mathcal{R}_0$) resulted in the collapse of the system in the case of $\Lambda=0$ [Fig.~\ref{fig:SIS1}(c)]. Finally, when $\mathcal{E}>\mathcal{E}_c^{+}$, the system evolves to the endemic state II(--) (stable point $s^*_-$) at any initial conditions [Fig.~\ref{fig:SIS3}(d)]. In this case, the stationary number density $s^*_-$ of active individuals is quite low, but still above zero. The corresponding stationary resource value $\rho^*$ is close to $s_\Lambda\rho_0$. When $\mathcal{E}$ is larger than $\mathcal{E}_c^{+}$, but still close to it, the evolution from the disease-free equilibrium (which is unstable in this case) to the endemic equilibrium $s^*_-$ takes quite a long time and passes through a quasi-stationary intermediate state, similar to the case of stair-like collapse described previously for the case $\Lambda=0$.

The above results indicate that positive $\Lambda$ serves as a mitigation factor to the collapse scenario considered in the previous subsection.

\subsection{Constant resource outflow ($\Lambda<0$)}

\begin{figure}[t]
\includegraphics[width=0.75\columnwidth]{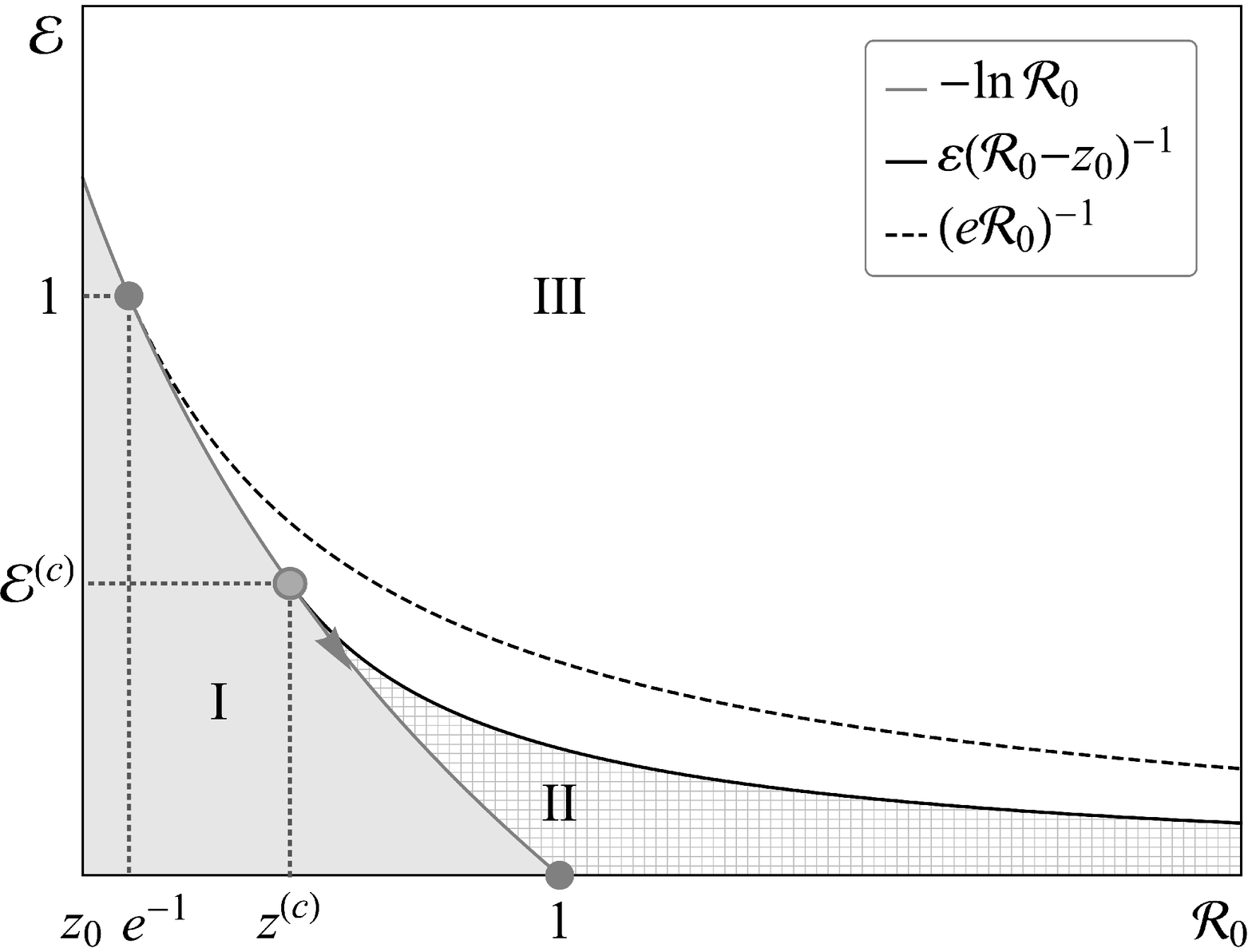}
\caption{\label{fig:phaseminus}Phase diagram for the coupled epidemic-resource system described by Eqs.~(\ref{eq:SIS}) with $\Lambda<0$. As $|s_\Lambda| = |\Lambda|/\,G$ increases, the phase boundary (solid black curve) between the endemic phase II and collapse phase III slides down across the logarithmic curve $\mathcal{E}=\mathcal{E}_e$ starting from its position at $s_\Lambda=0$ (dashed curve), until it reaches the point $\mathcal{E}=0$ at $|s_\Lambda|=\mathcal{R}_0^{-1}$. The triple point (point of contact between the phase boundaries) is located at $\mathcal{R}_0=z^{(c)}$, where $z^{(c)}$ is the position of the minimum of the transcendental function $(z-z_0)\ln z$ with $z=s^*\mathcal{R}_0$ and $z_0 = |s_\Lambda|\mathcal{R}_0$ [see Eq.~(\ref{eq:SIS_trans_Lambda})].}
\end{figure}

\begin{figure*}[!]
\includegraphics[width=\textwidth]{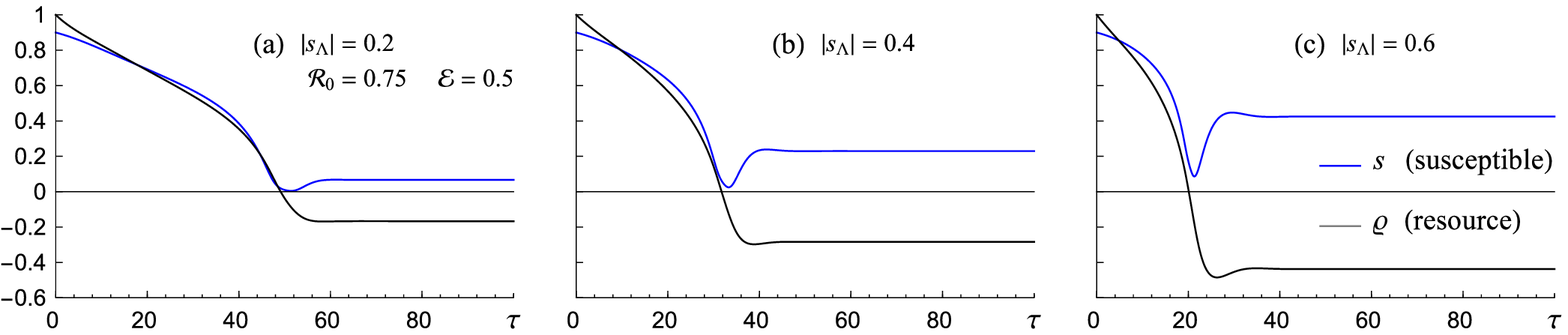}
\caption{\label{fig:SIS2}(Color online) The number density of susceptible (active) individuals and normalized resource function $\varrho=\rho/\rho_0$, with $\rho_0=(G-|\Lambda|)/\,\Gamma$, versus dimensionless time $\tau = \gamma_0\,t$ in the case of negative $\Lambda$ with all other parameters selected as in Fig.~\ref{fig:SIS1}(c). The collapse scenario is mitigated by means of debt (negative resource). The larger the infrastructure expenses ($|s_\Lambda| = |\Lambda|/\,G$), the greater is the debt required to finance external payments and the larger is the number of active (recovered) individuals.}
\end{figure*}

Negative $\Lambda$ means constant resource outflow from the system, e.g., in the form of infrastructure expenses, depreciation, rent, or interest payments. In this case, the stationary points of Eqs.~(\ref{eq:SIS}) are found similarly to the case $\Lambda>0$. The stationary solution to the equation for $\rho$ is given by formula (\ref{eq:rhostar_positive}), with \mbox{$s_\Lambda = -|\Lambda|/\,G<0$}. In this case, the parameter $|s_\Lambda|$ defines the minimum number of active agents required to secure external payments, e.g., to sustain some infrastructure. To keep the disease-free equilibrium resource $\rho_0$ positive, the infrastructure expenses should not exceed resource acquisition, so that we might restrict our attention to the case~$|\Lambda|<G$.

When the system leaves the disease-free equilibrium because of epidemic shock, it evolves to another equilibrium point, with the number density $s^{*}$ of active agents given by Eq.~(\ref{eq:SIS_trans_Lambda_positive}). While $s^{*}>|s_\Lambda|$, there are enough active agents to generate the necessary resource to secure all the external payments, so that we have $0<\rho^*<\rho_0$. This is the case of endemic equilibrium. The case $s^{*}=|s_\Lambda|$ means the complete depletion of resource, inasmuch as $\rho^*=0$. This signifies the onset of the collapse of the economic subsystem, despite the fact that there is still some number of active agents in the system. The corresponding critical activation energy for the transition from the endemic equilibrium to collapse is
$$
\mathcal{E}_c=\varepsilon\bigl(\mathcal{R}_0(1-|s_\Lambda|)\bigr)^{-1},
$$
where $\varepsilon$ is the absolute value of the local minimum of the transcendental function of $\mathcal{R}_0s^*$ in the left-hand side of Eq.~(\ref{eq:SIS_trans_Lambda_positive}). In particular, $\varepsilon=e^{-1}$ for $s_\Lambda=0$ [see Eq.~(\ref{eq:E_c})] and $\varepsilon=0$ for $|s_\Lambda|=\mathcal{R}_0^{-1}$. As $|s_\Lambda|$ increases from 0 to $\mathcal{R}_0^{-1}$, the triple point in the phase diagram shown in Fig.~\ref{fig:phaseminus} slides down across the logarithmic curve $\mathcal{E}=\mathcal{E}_e$ until it reaches the point $\mathcal{E}=0$.

When $s^{*}<|s_\Lambda|$, the stationary resource value $\rho^*$ becomes negative. In the statistical mechanics of money, debt is formally associated with negative money \cite{Yakovenko_2000,Yakovenko_rmp_2009,Chakra} or antimoney \cite{Antimoney_2014}. In this context, a negative resource formally means that there is no sufficient resource to secure external payments, and some of these payments need to be financed through a debt. In the next Section, we will show that allowance for a negative resource (debt) can be used to suggest one of the possible collapse mitigation strategies.

\section{Collapse mitigation}

Our system given by Eqs.~(\ref{eq:SIS}) consists of two subsystems: the economic one described by resource $\rho$ and the population one described by the number density $s$ of active individuals. Accordingly, it can be influenced either through the resource subsystem (e.g., using certain financial instruments) or through the population subsystem (e.g., introducing social regulations like quarantine or through vaccination). Here we consider several illustrative examples of the collapse mitigation strategies based on our model. We start from the strategies dealing with the resource subsystem and then proceed to the population subsystem.

\subsection{External subsidies or debt}

{\it Subsidies}.---This strategy pertains to the case of $\Lambda>0$, which was considered in Sect.~III.B. When $\Lambda$ is positive, which formally means some external source of subsidies or money inflow into the system, the hard collapse scenario considered in Sect.~III.A is mitigated. The larger the subsidy amount (parameter $\Lambda$), the higher is the endemic number density $s^*$ of active agents and the higher is the stationary resource value $\rho^*$. Both $s^*$ and $\rho^*$ stay positive at any $\mathcal{E}$ and $\mathcal{R}_0$, although $s^*$ can be very small at sufficiently high $\mathcal{E}$ [Fig.~\ref{fig:SIS3}(c,d)].
\smallskip

{\it Debt or negative resource}.---This strategy deals with the case $\Lambda<0$ and implies a change in the taxation policy through a certain modification of the model equations. When $\Lambda$ is negative, some part of resource is taken out from the system in the form of external payments like infrastructure expenses, depreciation, rent, interest payments, etc. On the one hand, this leads to the aggravation of the collapse scenario, when resource is depleted faster than in the case of $\Lambda=0$ (see Sect.~III.C). On the other hand, such a scenario allows for a different strategy that can mitigate the collapse, as demonstrated below.

We proceed from the standpoint adopted in the statistical mechanics of money implying that negative money can be associated with debt \cite{Yakovenko_2000,Yakovenko_rmp_2009,Chakra}. When an economic agent does not have enough money to pay his bills, he can borrow the required amount from an external reservoir (e.g., from a bank), and his balance becomes negative. When an agent with a negative balance earns some money, he uses this money to repay the debt until his balance becomes positive. The same ideology also applies to the concept of wealth, with negative wealth associated with debt \cite{Huang_2004}. In this work, resource $\rho$ is associated with average income per economic agent. By analogy to money and wealth, it can become negative if a group of economic agents described by Eqs.~(\ref{eq:SIS}) starts to live in debt (on the average), borrowing resource (money) from an external reservoir.

In terms of our equations, negative resource means that the term $\Gamma\rho$ changes its sign. This means that resource is no longer consumed but is collected in the form of debt from an external reservoir. In contrast to the case of positive $\Lambda$ (subsidies), such a resource inflow is not constant but is proportional to the number density of active agents [see Eq.~(\ref{eq:rhostar_positive})]. Such an economic model has some similarities with the the so-called negative income tax (NIT) \cite{NIT_Chicago,NIT_2020}. According to Ref.~\cite{NIT_MIT}, ``The negative income tax is a way to provide people below a certain income level with money. In contrast to a standard income tax, where people pay money to the government, people with low incomes would receive money back from the government.''

In the case of negative resource, relation (\ref{eq:gamma}) for the recovery rate should be modified appropriately to keep it finite. In the model considered in Ref.~\cite{Swiss_2015}, the domain of negative resource (budget) was excluded as nonphysical by putting $\gamma(\rho)\equiv0$ for $\rho\leqslant0$. In this work, we propose a different approach/strategy. Every strategy should be reflected in modified model equations. Here we rewrite relation~(\ref{eq:gamma}) in terms of the resource's absolute value, namely,
\begin{equation}
\gamma(\rho)\mapsto\gamma(|\rho|)= \gamma_0 \,\exp(-E/\,|\rho|),
\end{equation}
with asymptotic value $\gamma(\rho)=0$ at $\rho=0$. Such a proposition follows from the observation that the distributions of negative money (antimoney) and negative wealth follow the same exponential law as the positive money/wealth distributions \cite{Antimoney_2014,Huang_2004}.

With this modification, Eq.~(\ref{eq:SIS_trans_Lambda_positive}) for the stationary number density $s^{*}$ of active agents is rewritten as
\begin{equation}\label{eq:SIS_trans_Lambda}
\bigl|s^{*}-|s_\Lambda|\bigr|\,\ln\left(\mathcal{R}_0 s^{*}\right)=-\mathcal{E}\left(1-|s_\Lambda|\right).
\end{equation}
In contrast to the case $\Lambda=0$, this equation has one additional solution, $s_\rho^{*}<s_\Lambda$, that exists for any \mbox{$\mathcal{E}>0$}. It is always stable, and the corresponding stationary resource value is always negative. This solution is the direct counterpart to the collapse point $s_c^{*}=0$ existing in the case of $\Lambda=0$.

Thus, changing the taxation policy under the critical conditions ($\rho \rightarrow 0$) in the case of negative $\Lambda$ can serve as a mitigating factor to the collapse scenario, with the system stabilization achieved by means of debt (negative resource). Figure~\ref{fig:SIS2} demonstrates an example of such a mitigated collapse scenario for the same set of parameters as in Fig.~\ref{fig:SIS1}(c). The number density of active agents bounces from a horizontal axis close to $s=0$ and stabilizes at $s=s_\rho^{*}$, with resource passing through the zero point and stabilizing at the negative value given by relation~(\ref{eq:rhostar_positive}). The larger the parameter $|s_\Lambda|$, the greater is the debt required to finance external payments and the larger is the number of active (recovered) individuals.

\subsection{Quarantine scenario}

As mentioned in the Introduction, the use of one or another quarantine strategy against the epidemic can lead to a worsening of the general socioeconomic situation, at least for some social groups. Here we use our simple model to consider a speculative example demonstrating a possibility of different quarantine outcomes.

Quarantine measures are all aimed to reduce the contagion transmission rate, which is defined by the coefficient $\beta$ in our model given by Eqs.~(\ref{eq:SIS}). The objective of most quarantine strategies is to reduce the contact rate of individuals, their spatial density, to limit their social activity, to apply vaccination, etc. The transmission rate $\beta$ is determined, in particular, by the average frequency of collisions (contacts) between agents in population, depending on the mean local population density at normal conditions ($s\equiv1$ in the disease-free state). Here local density implies a characteristic density of individuals or mean distance between them in various social locations like transport, shop, work, etc., for a selected social group. Note that the density dependence of $\beta$ is generally non-monotonous: a lower density results in a lower collision probability, while a higher density causes a lower mobility of agents \cite{hao}.

Suppose that the reduction of $\beta$ is achieved by decreasing the local population density as a result of certain quarantine measures like social distancing, self-isolation, shortened workday, etc. These quarantine measures should also affect the resource acquisition rate $G$ (if it depends on local population density as well) and, as a result, the average income per agent. In this case, the use of such quarantine measures formally means that the rate constants $\beta$ and $G$ are renormalized, namely, $\beta\mapsto\beta'$ and $G\mapsto G'$, with $\beta'<\beta$ and $G'<G$.

\begin{figure}[!]
\includegraphics[width=.75\columnwidth]{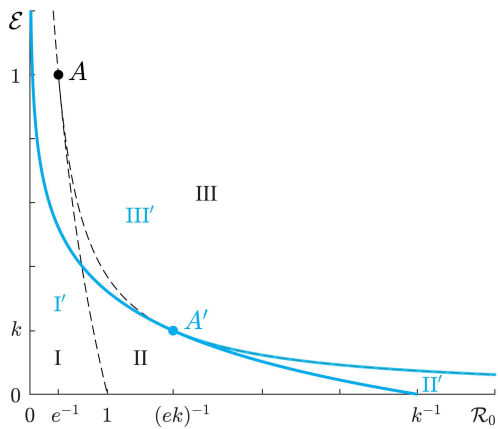}
\caption{\label{fig:quarantine}(Color online) Effect of quarantine transformation $\hat{Q}_k$ on the phase boundaries in the $(\mathcal{R}_0,\mathcal{E})$ plane for the case $\Lambda=0$. The dashed curves mark the boundaries of phase domains when there is no quarantine. The solid curves mark the transformed phase boundaries in the case of quarantine of strength $k=0.2$. The points $A=(e^{-1},\,1)$ and $A'=\bigl((ek)^{-1},\,k\bigr)$ are the contact points of the phase boundaries. Quarantine can have an ambiguous effect, depending on initial resource $\rho_0$ and activation resource (energy) $E$.}
\end{figure}

For simplicity, let us assume that $\beta'=k\beta$ and $G'=kG$, where the parameter $k$ ($0<k<1$) formally corresponds to the quarantine severity factor. The smaller the factor $k$, the stronger are the quarantine regulations. We also restrict our attention to the case $\Lambda = 0$.

The phase state of our system corresponds  to a point $(\mathcal{R}_0,\,\mathcal{E})$ on the phase diagram (Fig.~\ref{fig:phase}), with $\mathcal{R}_0=\beta /\gamma_0$, $\mathcal{E}=E\,\Gamma/G=E/\rho_0$, and $\rho_0$ being the initial resource. The above-described quarantine scenario formally corresponds to the scaling transformation
$$
\hat{Q}_k:\;\;(\mathcal{R}_0,\,\mathcal{E})\mapsto (\mathcal{R}'_0,\,\mathcal{E'})=(k\mathcal{R}_0,\,\mathcal{E}/k)
$$
that maps a given curve $\mathcal{E}=f(\mathcal{R}_0)$ into a curve $\mathcal{E}(k)=kf(k\mathcal{R}_0)$. The phase boundaries in Fig.~\ref{fig:phase} are formed by two curves,
$\mathcal{E}_c=(e\mathcal{R}_0)^{-1}$ and $\mathcal{E}_e=-\ln(\mathcal{R}_0)$,
with contact point $A=(e^{-1},\,1)$. Transformation $\hat{Q}_k$ maps them into the curves $\mathcal{E}_c(k)=\mathcal{E}_c$ and $\mathcal{E}_e(k)=-k\ln(k\mathcal{R}_0)$, with contact point $A'=\bigl((ek)^{-1},\,k\bigr)$ sliding down along the hyperbola $\mathcal{E}_c(\mathcal{R}_0)$ (Fig.~\ref{fig:quarantine}). This means that implementing the quarantine strategy $\hat{Q}_k$ deforms the phase boundaries and changes the positions of phase domains, namely, $\hat{Q}_k$:~(I,~II,~III) $\rightarrow$ (I$'$,~II$'$,~III$'$).

Figure \ref{fig:quarantine} demonstrates, on the one hand, that quarantine $\hat{Q}_k$ always leads to the expansion of the collapse domain (III $\subset$ III$'$), whose bottom (which is described by formula (\ref{eq:Eb})) goes down. On the other hand, quarantine results in the contraction of the endemic domain (II$'\subset$ II), whose bottom goes up and shifts. The disease-free domain partially expands at the same time. In other words, there exist phase points with relatively high $\mathcal{E}$ ($\mathcal{E}>k$) that, being initially located in the domain I or II, can appear in the collapse domain III$'$ after the quarantine transformation $\hat{Q}_k$. Alternatively, some points from domain II with relatively small $\mathcal{E}$ ($\mathcal{E}<k$) can appear in the disease-free domain I$'$. Note that quarantine of any severity would not prevent the system from collapsing, inasmuch as the collapse domain III does not get smaller at any $k<1$.

Thus, quarantine can have an ambiguous effect. For socioeconomic systems with small initial resource $\rho_0$ or high level of minimum resource consumption $E$, the above-discussed quarantine strategy can ultimately result in a collapse even if the system was initially in quite a controllable situation. On the contrary, quarantine always has a positive effect on systems with high $\rho_0$ or low $E$, such that it can even suppress the epidemic.

If a quarantine strategy involves only the reduction of the transmission rate $\beta$ and has no effect on the resource acquisition rate $G$ (e.g., in the case of vaccination campaign), it will not lead to a worsening of the general socioeconomic situation.

Here we considered the simplest example of the quarantine scenario supposing that the epidemic transmission and resource acquisition rates linearly depend on population density and using the trivial transformation of kinetic coefficients, $\beta'/\beta=G'/G$. More general forms of this transformation can be considered as well, namely, $\beta'\propto(k)^\mu\beta$, $G'\propto(k)^\nu G$, allowing for collective or coherent effects as well as non-trivial concentration dependencies. Such a consideration would require the use of a more general kinetic approach.

\section{Conclusion}

A simple model for the socioeconomic system that was considered here is based on the activation-type mechanism of the epidemic-resource coupling [see Eq.~(\ref{eq:gamma})]. Such a coupling mechanism naturally results in the collapsing effect opposite to the well-known thermal explosion. The activation parameter $E$ characterizes the minimum amount of the consumed resource needed for the survival of a community or a particular individual, therefore implying the existence of some ``energy'' barrier for their survival. Similar mechanisms are likely to be peculiar to other population systems as well. Note that a similar formal relationship between thermal explosion and a birth-death process was discussed in Ref.~\cite{Nicolis1983}, and its possible applicability to population dynamics and epidemiology was also mentioned in Ref.~\cite{NicPrig}.

In this work, we demonstrated that in the case of limited economic resource there exists a certain critical point at which the system collapses at any initial conditions and can no longer stabilize and return to the stable pre-epidemic or post-epidemic state. Such a scenario is possible even when the basic reproduction number $\mathcal{R}_0$ is smaller than unity, in contrast to the standard epidemic models, where the epidemic can spread only at $\mathcal{R}_0>1$ \cite{BookSpringer2015,BookSpringer2019}. Collapse is generally caused by the reduction of
the recovery rate $\gamma$. Besides the effect of the Arrhenius factor, which identifies the probability that the individual may recover or get help (e.g., a medical service), the recovery rate can also go down when the rate constant $\gamma_0$ in Eq.~(\ref{eq:gamma}) gets smaller. In particular, the parameter $\gamma_0$ formally takes into account the number of attempts the individual needs to make to get available response. In other words, the average time of delivery of medical services to the patients gets longer at smaller $\gamma_0$.

We considered several collapse mitigation strategies that can involve either financial instruments like subsidies and debt or social regulations like quarantine. We demonstrated that the system’s collapse can partially be mitigated by external subsidies meaning constant resource inflow from some external source or by means of debt interpreted as a negative resource. On the other hand, social regulations involving quarantine measures can have an ambiguous effect. When the initial equilibrium resource value is high enough and/or the minimum level of resource consumption is sufficiently low, the use of quarantine measures significantly improves the general socioeconomic situation. However, implementing strict quarantine measures in the case of low resource can finally lead to the collapse of the system being initially either in the endemic equilibrium or even in the disease-free equilibrium. These results provide a clear illustration to possible outcomes of epidemic-like systemic shocks.

In closing, let us make one remark. In the context of Refs.~\cite{Kusmartsev_2011,Xu_epl_2015,Tao_2010,Rashkovskiy_2019}, the crisis state of the financial market for a system of economic agents can be associated with a Bose condensate-like state at low market temperature. In this regard, it is interesting that our model given by Eqs.~(\ref{eq:SIS}) formally describes the dynamics of cooling of a system of agents due to contagion-induced transitions between two discrete inner states of agents (passive/active) characterized by two levels of $s=\{0,1\}$.

\acknowledgments{
O.K. was partially supported by a grant for research groups of young scientists from the National Academy of Science of Ukraine (Project No. 0120U100155). We thank Prof. B.I. Lev for fruitful discussions and the reviewers for their valuable comments.}


\begin{thebibliography}{99}

\bibitem{SystemicShock_2020}
W. Hynes, B. Trump, P. Love, and I. Linkov, Environ. Syst. Decis. {\bf 40}, 174 (2020).

\bibitem{Anderson_Lancet_2020-03}
R.M. Anderson, H. Heesterbeek, D. Klinkenberg, and T.D. Hollingsworth, Lancet {\bf 395}, 931 (2020).

\bibitem{Economics_2014}
C. Perrings, C. Castillo-Chavez, G. Chowell, P. Daszak, E.P. Fenichel, D. Finnoff, R.D. Horan, A.M. Kilpatrick, A.P. Kinzig, N.V. Kuminoff,
S. Levin, B. Morin, K.F. Smith, and M. Springborn, EcoHealth {\bf 11}, 464 (2014).

\bibitem{Optimization_SIS_2017}
H. Chen, G. Li, H. Zhang, and Z. Hou, Phys. Rev. E {\bf 96}, 012321 (2017).

\bibitem{Percolation_2017}
M. Schr\"{o}der, N.A.M. Ara\'{u}jo, D. Sornette, and J. Nagler, Phys. Rev. E {\bf 96}, 062302 (2017).

\bibitem{Bauch_Earn_2004}
C.T.~Bauch and D.J.D.~Earn, PNAS {\bf 101}(36), 13391 (2004).

\bibitem{Economics}
P.A. Samuelson and W.D. Nordhaus, \textit{Economics}, 19th ed. (McGraw-Hill, New York, 2009).

\bibitem{China_PR_2018}
W. Wang, Q.-H. Liu, J. Liang, Y. Hu, and T. Zhou, Phys. Rep. {\bf 820}, 1 (2018).

\bibitem{China_2018}
J. Jiang and T. Zhou, Sci. Rep. {\bf 8}, 1629 (2018); Physica A {\bf 508}, 414 (2018).

\bibitem{Swiss_PRL_2017}
L. B\"{o}ttcher, J. Nagler, and H.J. Herrmann, Phys. Rev. Lett. {\bf 118}, 088301 (2017).

\bibitem{Cascades_PRE_2015}
C.D. Brummitt and T. Kobayashi, Phys. Rev. E {\bf 91}, 062813 (2015).

\bibitem{Yakovenko_2000}
A. Dr\u{a}gulescu and V.M. Yakovenko, Eur. Phys. J. B {\bf 17}, 723 (2000); {\it ibid.} {\bf 20}, 585 (2001).

\bibitem{Yakovenko_rmp_2009}
V.M. Yakovenko and J.B. Rosser Jr., Rev. Mod. Phys. {\bf 81}(4), 1703 (2009).

\bibitem{Yakovenko_2010}
A. Banerjee and V.M. Yakovenko, New J. Phys. {\bf 12}, 075032 (2010).

\bibitem{Efthimiou_2016}
C.J. Efthimiou and A. Wearne, Eur. Phys. J. B {\bf 89}(3), 82 (2016).

\bibitem{Kusmartsev_2011}
F.V. Kusmartsev, Phys. Lett. A {\bf 375}, 966 (2011);
K.E. K\"{u}rten and F.V. Kusmartsev, EPL {\bf 93}, 28003 (2011).

\bibitem{Xu_epl_2015}
J. Xu, EPL {\bf 110}, 58002 (2015).

\bibitem{Tao_2010}
Y. Tao, Phys. Rev. E {\bf 82}, 036118 (2010).

\bibitem{Rashkovskiy_2019}
S.A. Rashkovskiy, Physica A 
{\bf 514}, 90 (2019).

\bibitem{hao}
H. Hu, K. Nigmatulina, and P. Eckhoff, Math. Biosci. {\bf 244}, 125 (2013).

\bibitem{Chatterjee_2007}
A. Chatterjee and B.K. Chakrabarti, Eur. Phys. J. B {\bf 60}, 135 (2007).

\bibitem{Chakra}
B.K. Chakrabarti, A. Chakraborti, S.R.~Chakravarty, and A.~Chatterjee, {\it Econophysics of Income and Wealth Distributions} (Cambridge Univ. Press, Cambridge, 2013).

\bibitem{BookSpringer2015}
M. Martcheva, {\it An Introduction to Mathematical Epidemeology} (Springer, New York, 2015).

\bibitem{BookSpringer2019}
F. Brauer, C. Castillo-Chavez, and Z. Feng, {\it Mathematical Models in Epidemeology} (Springer, New York, 2019).

\bibitem{Swiss_2015}
L. B\"{o}ttcher, O. Woolley-Meza, N.A.M. Ara\'{u}jo, H.J. Herrmann, and D. Helbing, Sci. Rep. {\bf 5}, 16571 (2015).

\bibitem{China_PRE_2019}
X. Chen, T. Zhou, L. Feng, J. Liang, F. Liljeros, S.~Havlin, and Y. Hu, Phys. Rev. E {\bf 100}, 032310 (2019).

\bibitem{RevModPhys_2015}
R. Pastor-Satorras, C. Castellano, P.V. Mieghem, and A.~Vespignani, Rev. Mod. Phys. {\bf 87}, 925 (2015).

\bibitem{Swiss_PRE_2016}
L. B\"{o}ttcher, O. Woolley-Meza, E. Goles, D. Helbing, and H.J. Herrmann, Phys. Rev. E {\bf 93}, 042315 (2016).

\bibitem{PRL_2020}
S. Moore and T. Rogers, Phys. Rev. Lett. {\bf 124}, 068301 (2020).

\bibitem{Laidler}
S. Glasstone, K.J. Laidler, and H. Eyring, \textit{The Theory of Rate Processes: The Kinetics of Chemical Reactions, Viscosity, Diffusion and Electrochemical Phenomena} (McGraw-Hill, New York, 1941); K.J. Laidler, \textit{Chemical Kinetics}, 3rd ed. (Pearson, 1987).

\bibitem{Stiller}
W. Stiller, \textit{Arrhenius Equation and Non-Equilibrium Kinetics: 100 Years Arrhenius Equation} (B.G. Teubner, Leipzig, 1989).

\bibitem{Yakovenko_2019}
Y. Tao, X. Wu, T. Zhou, W. Yan, Y. Huang, H. Yu, B.~Mondal, and V.M. Yakovenko, J. Econ. Interact. Coord. {\bf 14}, 345 (2019).

\bibitem{PRL_1966}
P.A. Sturrock, Phys. Rev. Lett. {\bf 16}, 270 (1966).

\bibitem{Zeldovich_Frank}
Ya.B.~Zel'dovich and D.A. Frank-Kamenetskii, Dokl. Akad. Nauk SSSR {\bf 19}, 693 (1938); D.A. Frank-Kamenetskii, {\it ibid.} {\bf 18}, 413 (1938).

\bibitem{Smirnov}
B.M. Smirnov, Sov. Phys. Usp. \textbf{34}, 526 (1991).

\bibitem{Novozh_2018}
V. Novozhilov, Sci. Rep. {\bf 8}, 4030 (2018).

\bibitem{Amado_2019}
A. Amado, J.V. Santana-Filho, P.R.A. Campos, and E.P.~Raposo, Eur. Phys. J. Plus {\bf 134}, 151 (2019).

\bibitem{Macroeconomic_2020}
M.S. Eichenbaum, S. Rebelo, and M. Trabandt, NBER Working Paper No. 26882 (2020).

\bibitem{Sugiarto_PRL_2017}
H.S. Sugiarto, J.S. Lansing, N.N. Chung, C.H. Lai, S.A.~Cheong, and L.Y. Chew, Phys. Rev. Lett. {\bf 118}, 208301 (2017).

\bibitem{Gardiner}
N.G. van Kampen, \textit{Stochastic Processes in Physics and Chemistry} (Elsevier, Amsterdam, 2006); C. Gardiner, \textit{Stochastic Methods: A Handbook for the Natural and Social Sciences} (Springer, Berlin, 2009); C. Gardiner and P. Zoller, \textit{Quantum Noise} (Springer, Berlin, 2004).

\bibitem{KermackMcKendrick}
W.O. Kermack and A.G. McKendrick, Proc. R. Soc. London A {\bf 115}, 700 (1927); {\it ibid.} {\bf 138}, 55 (1932).

\bibitem{Schlogl_1972}
F. Schl\"{o}gl, Z. Phys. {\bf 253}, 147 (1972).

\bibitem{Staircase}
P. Bak, Phys. Today {\bf 39}(12), 38 (1986).

\bibitem{Antimoney_2014}
M. Schmitt, A. Schacker, and D. Braun, New J. Phys. {\bf 16}, 033024 (2014).

\bibitem{Huang_2004}
D.-W. Huang, Phys. Rev. E {\bf 69}, 057103 (2004).

\bibitem{NIT_Chicago}
G. Burtless and J.A. Hausman, J. Political Econ. {\bf 86}, 1103 (1978).

\bibitem{NIT_2020}
K. Kroft, K. Kucko, E. Lehmann, and J. Schmieder, Am. Econ. J. Econ. Policy {\bf 12}, 254 (2020).

\bibitem{NIT_MIT}
R. Linke, \textit{Negative Income Tax, Explained} (MIT Sloan School online publication, 2018).

\bibitem{Nicolis1983}
F. Baras, G. Nicolis, M.M. Mansour, and J.W. Turner, J. Stat. Phys. {\bf 32}, 1 (1983).

\bibitem{NicPrig}
G. Nicolis and I. Prigogine, \textit{Exploring Complexity: An Introduction} (W.H. Freeman, New York, 1989).

\end{thebibliography}
\end{document}